\documentclass[final]{aipproc}
\layoutstyle{8x11double}

\begin{document}

\title{\fontsize{16pt}{13pt}\selectfont The Initial State of Students Taking an Introductory Physics MOOC}

\classification{01.40.Fk,01.40.G-,01.40.gb}
\keywords{mechanics, assessment, participation, mooc, online}

\author{John M. Aiken}{
  address={Department of Physics and Astronomy, Georgia State University, Atlanta, GA, 30303}
  }

\author{Shih-Yin Lin}{
  address={School of Physics, Georgia Institute of Technology, Atlanta, GA, 30332}
  }

\author{Scott S. Douglas}{
  address={School of Physics, Georgia Institute of Technology, Atlanta, GA, 30332}
  }

\author{Edwin F. Greco}{
  address={School of Physics, Georgia Institute of Technology, Atlanta, GA, 30332}
  }

\author{Brian D. Thoms}{
  address={Department of Physics and Astronomy, Georgia State University, Atlanta, GA, 30303}
  }

\author{Michael F. Schatz}{
  address={School of Physics, Georgia Institute of Technology, Atlanta, GA, 30332}
  }

\author{Marcos D. Caballero}{
  address={Department of Physics and Astronomy, Michigan State University, East Lansing, MI 48824}
  }

\begin{abstract}
As part of a larger research project into massively open online courses (MOOCs), we have investigated student background, as well as student participation in a physics MOOC with a laboratory component.
Students completed a demographic survey and the Force and Motion Conceptual Evaluation at the beginning of the course.
While the course was still actively running, we tracked student participation over the first five weeks of the eleven-week course.
\end{abstract}

\maketitle

\section{\label{sec:intro}Introduction \& Background}

The development of research-based instructional transformations \cite{McDermott:1999tz,Meltzer:2012eg} is being outpaced by technology-driven pedagogical changes \cite{NYT2012}. 
As the rise of various technological advancements helped drive the development of distance education and university extension courses \cite{larreamendy2006going}, the delivery of massively open online courses (MOOCs) is becoming increasingly attractive as certain web technologies become more sophisticated and scalable. 
In the last two years, a large number of prestigious colleges and universities have entered into agreements with MOOC providers in order to reach more students, reduce instructional costs, and generate additional revenue.
Yet, the outcomes of learning in these online communities have only recently been investigated in limited contexts \cite{breslow2013studying}.
Furthermore, the study of how students engage in these communities is less understood.
By systematically investigating the development, implementation, and outcomes of MOOCs, we can help inform the community of colleges and universities of the benefits and challenges of these new web-based learning communities.

At Georgia Tech, we have begun investigating an introductory physics MOOC (Your World is Your Lab, YWYL) offered through the Coursera platform.
This course, unlike any other physics MOOC offered to date, is built from the on-campus implementation of a large-enrollment introductory mechanics course \cite{NNews}.
YWYL engages students in activities that are closely aligned with the on-campus experience including lectures, clicker questions, homework, exams, and laboratories.
In fact, students enrolled in a special Georgia Tech physics section will be recieving course credit for successfully completing this MOOC.
As of the writing of this paper, YWYL is in its sixth of eleven instructional weeks.

As part of the research conducted around YWYL, we have investigated the initial state of MOOC students including their background and pre-instruction content knowledge. 
We have also begun to investigate student participation in different aspects of the course. 
Students often pick and choose certain aspects with which to engage because the nature of these online environments and the ease of leaving these courses.
The laboratory component of YWYL represents a significant departure from the typical MOOC offering.
MOOCs have typically focused almost exclusively on video lectures and homework assignments.
Furthermore, the laboratory component is a substantial investment of the students' time. 

This paper serves the dual purpose of describing one model for implementing laboratory activities into a MOOC and of documenting student background and participation in a physics MOOC modeled after an on-campus implementation.
Key questions we attempt to answer include: who is taking this MOOC? and with what aspects of the course are they engaging and in what numbers? 
We also discuss what additional research is planned after course completion.

\vspace*{-14pt}
\section{\label{sec:mooc}A MOOC with Labs}

Your World is Your Lab (YWYL) is an eleven-week physics course that closely follows the large enrollment, calculus-based mechanics course offered at Georgia Tech. 
YWYL was designed to cover all topics presented in the on-campus introductory course including Newton's laws, work and energy, and rotational motion.  
For each of these topics, YWYL students have the opportunity to engage in activities borrowed from the on-campus course: lectures, textbook readings, homework assignments, laboratories, and exams.
Laboratories are central to YWYL, accounting for 65\% of a student's score in the course. 
While YWYL students are unable to participate in some key social aspects of the on-campus experience, online discussion forums are available to provide a proximal experience.
Teaching assistants moderate these forums and are available to help students through any difficulties with course material.

Content developed for the course was borrowed directly or modeled exclusively from course content for the on-campus implementation.
Lecture videos were developed from lectures delivered on campus, but we attempted to further arouse student interest in physics using animated lecture videos such as this one describing core ideas and practices, \url{http://goo.gl/87qIcB}. 
Most lecture videos were designed to engage students in learning physics content, but additional videos were developed to discuss important scientific practices like making models, using computation, and comparing models and data. 
Quizzes embedded in lecture videos simulate clicker questions.
Homework and exam questions were ported directly into the Coursera system.

The laboratories for YWYL were taken directly from the most recently delivered on-campus mechanics course. 
Unlike a traditional lab experience in which students work in a designated laboratory with special equipment, students enrolled in YWYL complete at-home labs using minimal equipment.
Starting the second week of class, students observe motion in their own environment and build computational models to explore the observed motion every other week (five total labs). 
In particular, students capture video of real-world objects using smartphones and then analyze the motion of these objects using open-source motion tracking software \cite{tracker}. 
Following their video analysis, students create computational models of the phenomena, using VPython \cite{VPython}, that they compare to their observations.  
Finally, students create and submit $\sim$5-minute long video reports that describe the experiment, the computational models, and how the results of each are related. 
To inform students how to complete these tasks, several lecture videos were produced to introduce students to the installation and use of the required software. 

If a student chooses to participate in all aspects of the course, a typical instructional week requires 11--14 hours of her time, which is comparable to the on-campus experience.
Each week, students are asked to watch seven to twelve lecture videos that typically last 5--15 minutes. 
Suggested reading assignments drawn from the optional textbook \cite{mandi1} are listed on the course website next to each lecture video. 
In addition to this instruction, students are also tasked with practicing what they are learning by completing two to three homework assignments every week.
Each homework assignment takes students 40--60 minutes to complete.
The five laboratory assignments follow a two-week cycle: 
In the first week, students perform the lab activity (video analysis and computer modeling), and then submit a video lab report. 
In the second week, each student is randomly assigned five reports from their peers, and asked to evaluate their peers' lab reports using an instructor-developed rubric.
As with most other MOOCs, students are also encouraged to participate in online discussions with the instructor, the TAs, and other students. 
Table \ref{tab:act} summarizes the major components of YWYL.

\begin{table}
\begin{tabular}{lll}\hline
{\bf Activity} & {\bf Frequency} & {\bf Estimated Time}\\\hline\hline
Video Lectures & 7--12 per week & 5--15 min. each\\
Homework & 2-3 per week & 40--60 min. each\\
Lab: Assignment & 1 every 2 weeks & 3--5 hours each\\
Lab: Evaluation & 1 every 2 weeks & 40--60 min. each\\
Discussion Forum & encouraged & variable\\\hline\hline
\end{tabular}
\caption{Students are asked to spend 11--14 hours working with core components of YWYL. For students who are fully engaging in the course, the majority of their time is devoted to homework sets and laboratory activities.\label{tab:act}}
\end{table}

While in principle YWYL could mirror most of the on-campus experience, a number of issues limit the use of research-based pedagogies, particularly those that leverage social learning, and, thus we are restricted in how closely the course can resemble the on-campus experience.
While the lecture videos are entertaining, we lack a technical solution for deploying clicker questions effectively.
Videos pause for students to reflect on and then answer clicker questions, but students can simply skip them altogether. 
Moreover, the lack of any common meeting environment limits any form of Peer Instruction \cite{mazurpeer}.
Students on campus participate in weekly presentation sessions where they present preliminary results from their experiments, which are critiqued by the other students and the classroom instructor.
These sessions aim to teach students how to develop appropriate video presentations and how to communicate effectively.
There is no formal mechanism for these sessions with online students.
However, some students have begun to use the forums for this purpose.

\vspace*{-14pt}
\section{\label{sec:students}Results \& Discussion}

As a first step to understanding these environments, we characterized student background using a demographic survey and the Force and Motion Conceptual Evaluation (FMCE) \cite{thornton1998assessing}, and investigated participation rates in various course components.
The data collected by Coursera includes every action by every student on every page, which produces an enormous amount of data for any one course.
In future work, we will delve deeply into the mountain of data collected in this course.

One of the central principles of massively open online courses is providing educational resources around the world for free. 
For educational researchers, this central principle is connected to issues of access and equity. 
While in principle, the openness of YWYL should make it accessible to people everywhere, those who live in developed countries were far more likely to enroll in YWYL.
High-speed internet with no restrictions is a necessity for accessing content in YWYL. 
A number of students living in countries with poor or restricted access to the web reported difficulties engaging in the course.
Such restrictions are both economic and cultural and are likely to restrict participation from certain regions of the world.

\vspace*{-14pt}
\subsection{Student Background}

In the first few weeks of the course, students were asked to complete a demographic survey that included questions about their age, sex, income, educational levels, location, and physics background.
Of the 18829 students registered for the course, 3092 students (16\%) completed the survey.
While the fraction of students completing the survey is low, we demonstrate below why we believe it is representative of students who are continuing to engage with the course.
It is from this subset that we draw characterizations of students taking YWYL.

Nearly half of YWYL students (44\%) are from the US and Canada.
European and Asian students constitute almost one-third of participants (31\%).
One-eighth of students (12\%) are distributed among Latin America (9\%), Africa (2\%), and Oceania (1\%).
The remaining students (13\%) chose not to report their location. 
Based on survey responses, students tend to be male (66\%) and younger; nearly 70\% are under 35.
YWYL students also tend to be educated; 85\% have earned at least a high school diploma and 59\% hold a college degree. 
These college degrees represent all fields, but are concentrated in traditional STEM and STEM education disciplines (67\%).
The course attracts students who have had some experience with physics. 
The majority of students (79\%) have taken at least high school physics, though many (46\%) took additional physics courses in college. 
Because YWYL requires students to complete at-home laboratories that include computational modeling, we investigated students' time commitment and computational background.
Few students reported they would spend the suggested 11--14 hours on the course; 83\% of students planned to spend less than nine hours a week on the course.
As we expected, most students (77\%) reported to have little to no programming experience.

\begin{figure}[t]
\includegraphics[width=0.90\columnwidth, clip, trim=5mm 5mm 5mm 5mm]{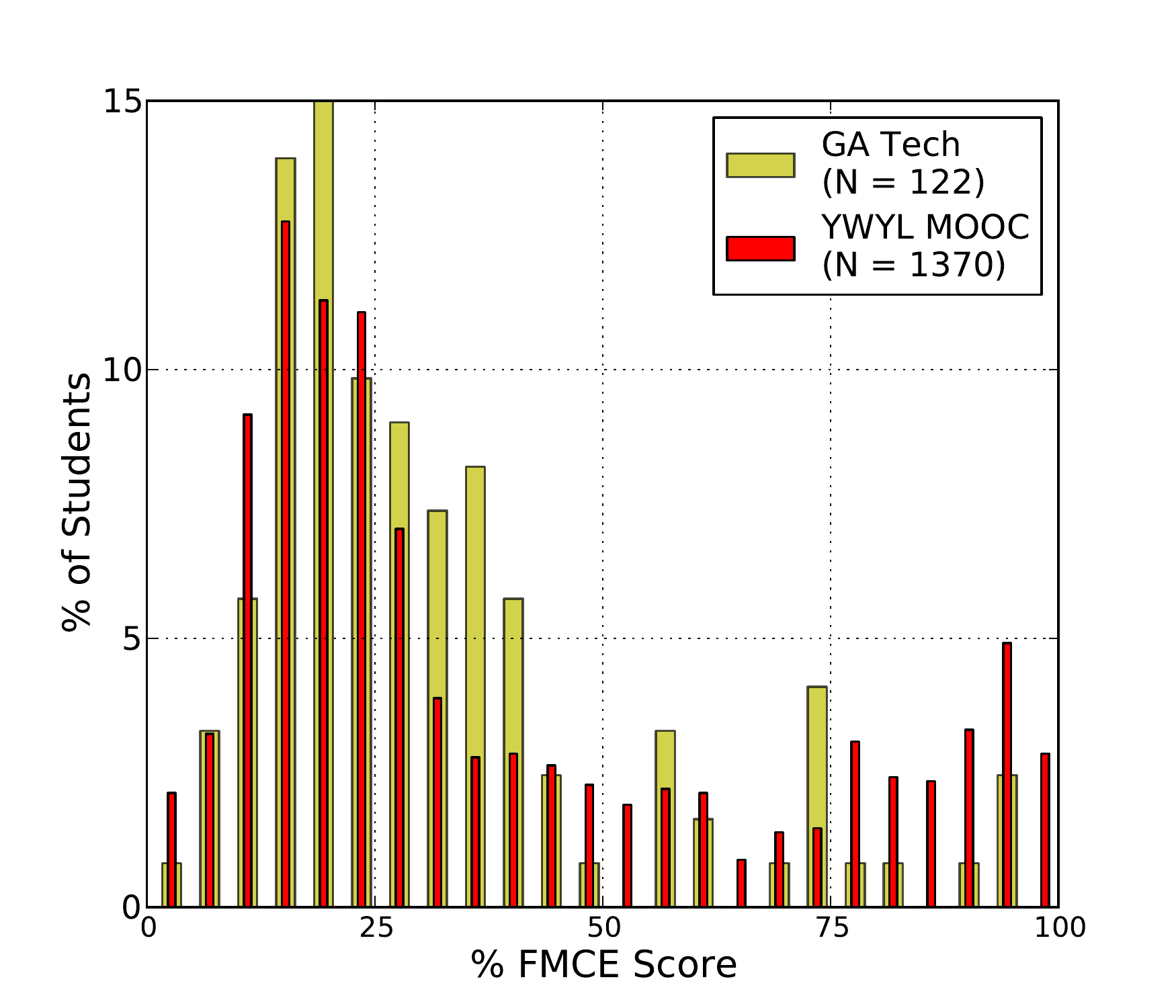}
\caption{Pre-test FMCE scores earned by YWYL students (N=1370). 
Representative FMCE pretest data collected at Georgia Tech are also presented here.
}
\label{fig:fmce}
\end{figure}

To gauge students' incoming conceptual knowledge, we collected student responses to the Force and Motion Conceptual Evaluation \cite{thornton1998assessing} in the first week of the course.
On the pre-test, YWYL students earned an average score of 39.9 $\pm$ 0.8\% (median 27.7\%). 
 Fig.\ \ref{fig:fmce} shows the distribution of scores, which appears bimodal. 
Nearly a third of students (30\%) earned pre-test scores above 50\%, which is uncharacteristic of introductory mechanics courses (e.g., \citet{thornton2009comparing}).
The top quintile earned scores above 74\% with an average score of 90.1\% $\pm$ 7.8\% (median, 93.6\%).

Students appearing in the top quintile of the FMCE score distribution are more well-educated than the general population; 70\% hold a college degree
In this top quintile, the most commonly degree held was at the Bachelor's level (29.6\%) followed closely by the Master's level (29.2\%).
Moreover, those degrees were more likely to be earned in STEM or STEM education disciplines (71\%).
Over half of these students took physics at the high school level (58\%) or college introductory level (48\%).
Over one-third took advanced undergraduate courses (35\%) and one quarter (23\%) took graduate physics courses.
Hence, this subpopulation is not representative of the typical introductory physics population.

\vspace*{-14pt}
\subsection{Student Participation}

To date, students have completed the first five weeks of the eleven-week course. 
At launch, 14474 students were registered for the course.
Because Coursera allows students to register at any time during an active course, students have continued to register in the course at a rate of $\sim$900 per week.
As of the sixth week of the class, 18829 students are registered.

\begin{figure}[t]
\includegraphics[width=0.90\columnwidth, clip, trim=8mm 10mm 10mm 10mm]{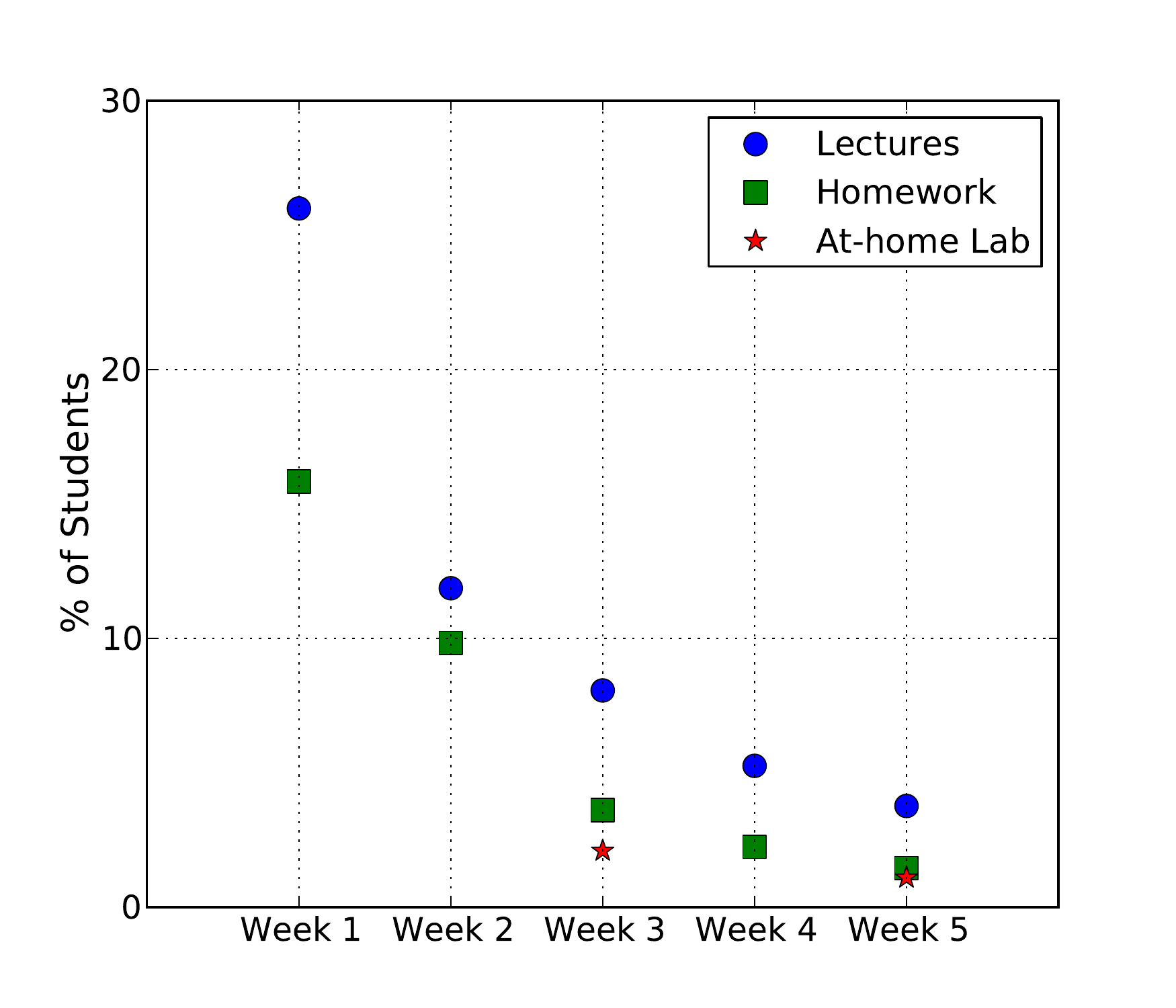}
\caption{Percentage of students participating in YWYL course activities averaged over each of the  first five weeks of the course (N=18829). Coursera allows students to register for the course at any time, so the overall number of students registered increases from week-to-week. The characteristic dropoff in student participation has been observed in other courses (e.g., \citet{breslow2013studying}).}
\label{fig:activity}
\end{figure}

However, as illustrated by Fig.\ \ref{fig:activity}, student participation is much lower than registration suggests.
Of all course activities (Table \ref{tab:act}), lecture videos require the least investment for students.
In the first week, the twelve lecture videos were viewed by an average of one quarter (26\%) of registered students.
Roughly 40\% of the students watched the first video, but less than one-fifth (17\%) watched the twelfth video.
By the fifth week, 4\% on average are watching lecture videos.
Attenuation has been observed in other courses (e.g., \citet{breslow2013studying}), but YWYL appears to lose students at a comparatively higher rate.

For activities that require more of the student's time such as homework and laboratories, we observe a strong attenuation rate and fewer students participate in these activities overall.
The two homework assignments given in the first week were attempted, on average, by one-sixth of the students (15\%).
By the fifth week, the number of students doing homework decreased by a factor of seven (2\%).
Even fewer students submit the laboratories; 2\% performed lab 1 and 1\% completed lab 2.

We have found one clear distinguishing demographic feature for students who are continuing to participate in the course, which we define as completing both laboratories.
On the initial demographic survey, students who completed both of the first two labs reported that they planned to spend slightly more time (9 or more hours) with the course than students who did not complete the first two labs ($\chi^2/\nu$ = 9.29, p $\ll$ 0.05) \cite{bevington}.
There appear to be no other distinguishing demographic characteristics about students who completed both labs.

The laboratory is the central aspect of YWYL, an idea communicated clearly in the first few lecture videos and on the syllabus. 
The laboratory represents 65\% of the credit towards certification; hence, no student can earn certification without completing most of the laboratories.
It is likely that the high attenuation rate is a result of the heavy emphasis on laboratories.
Upon course completion, students will be offered an exit survey asking them to describe what issues kept them engaged or drove them from the course.

\vspace*{-14pt}
\section{\label{sec:closing}Concluding Remarks}

Your World is Your Lab is an experiment into what is possible with the MOOC environment. 
While it remains to be seen how many students will complete the course, what they will have learned, and what factors contributed to their success, we believe that YWYL demonstrates an upper limit on what MOOC students will engage in.
This claim is based, in part, on the time commitment for the YWYL course.
Similar to the on-campus offering, we estimated students should spend roughly 11--14 hours per week on the course.
However, less than one-fifth of students (17\%) initially expected to spend more than nine hours working with course material.
This expectation is more in-line with the average weekly time commitment (6.6 hours/week) recommended by physics instructors of currently offered Coursera physics courses.
Only one out of 20 of the currently offered physics courses (not including YWYL) recommends spending more than 10 hours per week on their course.
It is also likely that the heavy emphasis on laboratory activities in YWYL has driven many students from the course, such that lowering the grade impact of the laboratories in future offerings might lead to lower attenuation and broader participation.

\vspace*{-14pt}
\begin{theacknowledgments}
The authors acknowledge our animators and coders who helped create much of the content for YWYL. 
We also thank Ron Thorton for the use of the FMCE.
This work was supported by the Bill \& Melinda Gates Foundation.
\end{theacknowledgments}

\vspace*{-14pt}
\bibliographystyle{aipproc}
\bibliography{mooc}

\end{document}